# Electrical Spin Pumping of Quantum Dots at Room Temperature

C. H. Li, G. Kioseoglou, O. M. J. van 't Erve, M. E. Ware, D. Gammon, R. M. Stroud, and B. T. Jonker

*Naval Research Laboratory*, Washington, DC 20375

R. Mallory, M. Yasar, and A. Petrou

State University of New York at Buffalo, Buffalo, NY 14260

ABSTRACT

We report electrical control of the spin polarization of InAs/GaAs self-assembled quantum dots (QDs) at room temperature. This is achieved by electrical injection of spin-polarized electrons from an Fe Schottky contact. The circular polarization of the QD electroluminescence shows that a 5% electron spin polarization is obtained in the InAs QDs at 300 K, which is remarkably insensitive to temperature. This is attributed to suppression of the spin relaxation mechanisms in the QDs due to reduced dimensionality. These results demonstrate that practical regimes of spin-based operation are clearly attainable in solid state semiconductor devices.

Semiconductor quantum dots (QDs) are nanostructures that have been described as "artificial atoms," since they exhibit discrete rather than continuous energy levels[1,2]. The lack of available energy states inhibits both elastic spin flip and inelastic phonon scattering mechanisms, resulting in long spin relaxation times[3,4,5]. This property is a critical requirement for the operation of various spin-based devices proposed, and makes QDs very attractive for the design and development of semiconductor spintronic devices and certain implementations of quantum information technology. Therefore, electrical injection of spin-polarized carriers into QDs is an important requirement that would impact a spectrum of future electronic and optical device applications.

Spin polarization of QDs has indeed been demonstrated, albeit at cryogenic temperatures. Electrical injection of holes from a ferromagnetic GaMnAs layer into self-assembled InAs/GaAs QDs[6,7], and optical pumping of a paramagnetic BeMnZnSe layer with transfer of electrons into CdSe/ZnSe QDs[8], have resulted in partial circular polarization of the QD electroluminescence as the spin polarized carriers radiatively recombined. While low temperature operation may suffice for certain niche applications, a broader technology demands room temperature operation.

We demonstrate here the ability to electrically control the spin polarization of self-assembled InAs/GaAs QDs embedded in an AlGaAs/GaAs quantum well *at room temperature*. We achieve an electron spin polarization in the InAs QDs of 5% at 300 K via injection of spin-polarized electrons from a reverse-biased Fe Schottky contact. This polarization is independent of temperature over a broad range (80-300K). These results demonstrate that practical regimes of spin-based operation are clearly attainable in future QD-based semiconductor devices.

The samples were grown by molecular beam epitaxy (MBE) in interconnected growth chambers[9], and consisted of a quantum well (QW) structure of 830 Å $n$-$Al_{0.1}Ga_{0.9}As$ / 400 Å undoped GaAs / 500 Å $p$-$Al_{0.3}Ga_{0.7}As$ / $p$-GaAs buffer layer on a $p$-GaAs(001) substrate. The QD layer was embedded in the middle of the undoped GaAs QW region. The top 150 Å of $n$-type $Al_{0.1}Ga_{0.9}As$ was highly doped ($n=1\times10^{19}$ cm$^{-3}$) to form the Schottky tunnel contact[10,11]. A 100 Å thick Fe(001) film was grown in a separate MBE chamber with the substrate at <5 °C to minimize potential intermixing at the Fe/$Al_{0.1}Ga_{0.9}As$ interface. Additional details of the growth are described elsewhere[9,10,11]. A schematic flat band diagram of the sample structure is shown in Fig. 1.

The quantum dots were formed by Stranski-Krastanov strain driven self-assembly. An indium flush procedure was used to obtain very precise control of the height of the dots at ~35 Å[12,13], but provides less control over the in-plane diameter. An atomic force microscopy image of a typical dot distribution is shown in Fig. 2(a), and reveals a rather broad size distribution with dot diameters ranging from 10 – 25 nm. A cross-sectional transmission electron microscopy (TEM) image of one of the spin-LEDs studied here (Fig. 2b) shows two of the dots in the QD layer. The coherence of the lattice fringes across the InAs dots and GaAs QW demonstrates the high quality growth.

The samples were processed to form surface emitting LEDs using standard photolithography and chemical etching techniques. The light emitted along the surface normal was analyzed and spectroscopically resolved by a 1/2 meter single pass spectrometer equipped with a liquid nitrogen cooled InGaAs array detector.

The circular polarization of the surface-emitted electroluminescence (EL) is a direct and quantitative measure of the electron spin polarization due to the quantum selection rules which govern the radiative recombination process in semiconductors[14]. For a QW-based spin-LED, the spin polarization of the electron population in the QW, $P_{QW}$, is equal to the circular polarization of the surface-emitted EL, $P_{circ}$[9,14], where $P_{circ}$ is defined as the normalized difference in intensities between the positive (σ+) and negative (σ-) helicity components: $P_{circ} = [I(σ+) – I(σ-)] / [I(σ+) + I(σ-)]$. The situation is more complicated for EL from QDs due to uncertainties in the dots' shape, since they are neither perfectly spherical nor round. Recent calculations, however, have shown that measuring the circular polarization of the QD EL along the surface normal (growth direction) with the carrier spin parallel to the photon momentum will give an accurate measure of the electron spin polarization of the QD, $P_{QD} = P_{circ}$, since this measurement geometry is not very sensitive to details of the QD shape[15]. This is the measurement geometry employed here.

Figure 3 shows the EL spectra at 80, 120, 200 and 300 K for a magnetic field of 3T applied along the surface normal. The zero field spectra at 300 K appear in the inset. All the spectra exhibit a predominant feature due to recombination of the ground state exciton (often referred to as the "*s*-shell" exciton) in the QDs. This fe ature is centered at ~ 1.21 eV at 80 K, and shifts to ~ 1.20, 1.16 eV and 1.13 eV at 120, 200 and 300 K, respectively, due to the decrease of the QD band gap with increasing temperature. The full width at half maximum (FWHM) of this feature reflects the size distribution of the QD ensemble, and is 62, 55, 45 and 48 meV at 80, 120, 200 and 300 K. This temperature dependence of the FHWM (minimum at an intermediate temperature between 100-200 K) is consistent with results published previously[1,16]. A slight shoulder on the higher energy side of the s-shell emission (labeled "p") is observed at 300 K, and is attributed to the QD excited state (p-shell) which is thermally populated only at higher temperatures.

At zero field (inset), the σ+ and σ– components are coincident ($P_{circ}$ = 0) because the Fe easy magnetization axis (and electron spin orientation) lies in-plane[14] and perpendicular to the QD hole spin orientation. A magnetic field applied along the surface normal rotates the Fe magnetization out-of-plane so that the quantum selection rule analysis described earlier can be applied to quantify the QD spin polarization. The EL spectra at 3T (sufficient to saturate the Fe magnetization along the surface normal) then exhibit a significant difference in intensity between the σ+ and σ- components, as seen in Fig 3, indicating a net electron spin polarization in the InAs QDs.

The magnetic field dependence of $P_{circ} = P_{QD}$ is shown in Fig. 4 for temperatures of 120, 200 and 300 K. Note that the $P_{QD}$ mirrors the hard axis magnetization of the Fe film (dashed line) obtained by independent superconducting quantum interference device (SQUID) magnetometry measurements, confirming that $P_{QD}$ is due to electrical injection from the Fe contact. $P_{QD}$ saturates at a value of about 5% when the Fe magnetization is fully out-of-plane. The positive sign of the polarization indicates injection of majority spin electrons from the Fe contact, consistent with earlier work with Fe Schottky contacts[10,11] and Fe/Al$_2$O$_3$ based tunnel junctions[17,18].

Control experiments were performed to rule out spurious effects. LED structures fabricated with the Fe contact removed showed little circular polarization for the temperature range discussed here. In addition, possible contributions to the measured $P_{circ}$ from Faraday rotation as the emitted light passes through the Fe film (magnetic dichroism) were assessed as described previously.[10,11] These background contributions are shown as the open symbols in Fig. 4, and are ≤1%[11], significantly smaller than the effect measured here and attributed to spin-polarized electron injection.

The QD spin polarization shows remarkably little dependence on temperature, as shown in the inset to Fig 4. This behavior is in marked contrast to the temperature dependence of the spin polarization in the QW case[10,19], where a significant reduction in polarization is observed over this same temperature range. A variety of mechanisms contribute to electron spin relaxation in III-V semiconductors, with GaAs being the most well studied[3-5,20,21,22,23]. At higher temperatures (>50 K), the D'yakonov-Perel' (DP) process typically dominates[3-5,20-23], in which a free electron moving through the lattice experiences an effective magnetic field due to spin-orbit coupling, hence randomizing the spin state. For a QW, this gives rise to a spin lifetime whose temperature dependence is typically given by $T^{-0.5}$.[14] A higher degree of spatial confinement (as in a QD) prevents electron motion in the lattice and the accompanying DP spin scattering[3-5], which should result in much longer spin lifetimes. In addition, the lack of available energy states due to the discrete nature of the QD density of states also inhibits both elastic spin flip and inelastic phonon scattering. If the resulting QD spin lifetime is greater than the radiative recombination lifetime at a given temperature (as found experimentally at low temperatures[3]), a fairly temperature independent behavior can be expected. The weak temperature dependence observed for $P_{QD}$ in figure 4 is consistent with suppression of the DP spin relaxation. Our results demonstrate one of the practical advantages offered by such zero-dimensional structures and their potential in the implementation of future spin-based semiconductor devices.

The QD spin polarization is smaller than that observed at low temperatures (5 K) from quantum well-based spin-LEDs, which ranges from 32-40%.[10,11,18] This is probably due to roughness at the Fe/AlGaAs interface which we observe with TEM, and attribute to perturbation of the AlGaAs surface as the QDs are incorporated into the structure. Defect structure at the spin injecting interface is known to reduce spin injection efficiency[24]. This may be corrected by modifying growth procedures in the future. Furthermore, the EL spectra represent an ensemble average of emission from millions of QDs. It is likely that electrical spin injection and analysis of individual dots will reveal significantly higher spin polarization and more detailed insight.

In summary, we have demonstrated electrical control of the spin polarization of self-assembled InAs/GaAs QDs *at room temperature* by electrical injection from a ferromagnetic Fe Schottky tunnel contact. This spin polarization is remarkably insensitive to temperature and persists to 300 K, consistent with suppression of the significant spin relaxation mechanisms due to the reduced dimensionality. Our results together with improved control of dot formation during growth to optimize dot size and location promise to enable strong spin-based effects with potential for single dot spin manipulation at room temperature which can be exploited for spintronic devices.

This work was supported by ONR (N0001404WX20052), the DARPA SpinS program (K920/00), and core programs at the Naval Research Laboratory.

**FIGURE CAPTIONS**

**Figure 1**      Flat band diagram of the QD spin-LED.

**Figure 2**      (a) A representative AFM image (0.6x0.4 µm$^2$) of the QDs. (b) A TEM cross-sectional image of the active area of one of the spin-LEDs studied here, showing two QDs within the GaAs QW.

**Figure 3**      EL spectra at 80, 120, 200 and 300 K for a magnetic field of 3 T, analyzed for positive (σ+) and negative (σ-) helicity, and (inset) at 300 K and zero field. The sample bias is 2.65 V.

**Figure 4**      The magnetic field dependence of the QD electron spin polarization, $P_{QD}$, for temperatures of 120, 200 and 300 K tracks the Fe out-of-plane magnetization as determined by SQUID magnetometry (dashed line). The open symbols show the field dependence of background contributions from control samples. Inset: Temperature dependence of the QD spin polarization.

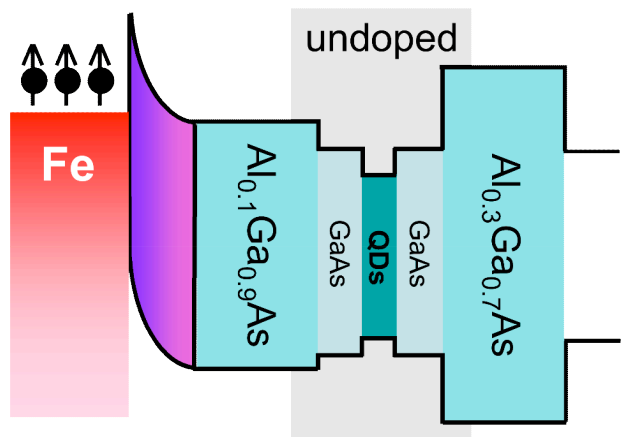

Li *et al*. Fig. 1

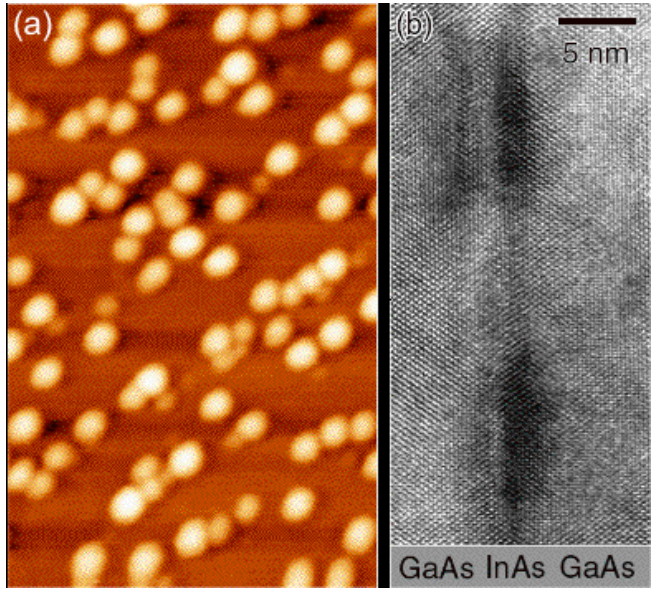

Li *et al*. Fig. 2

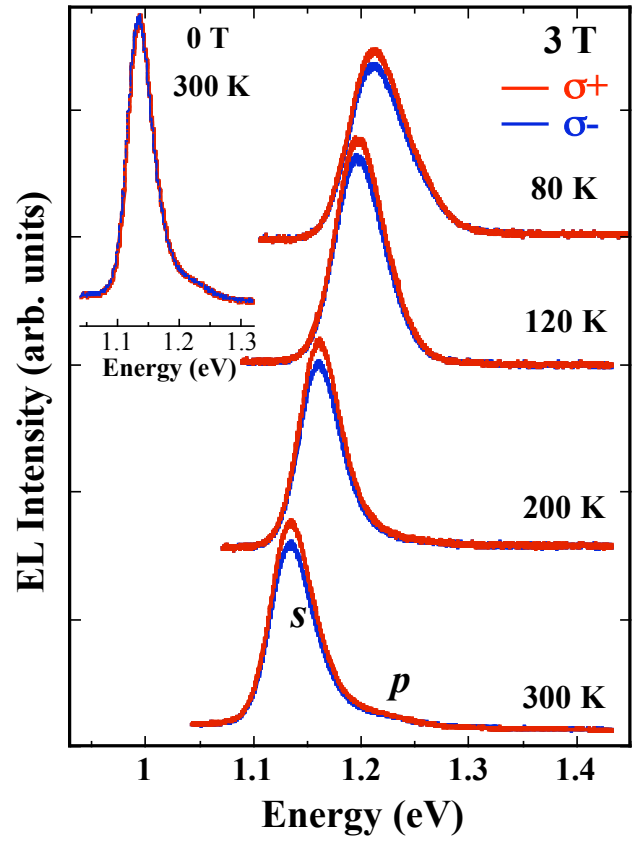

Li *et al*. Fig. 3

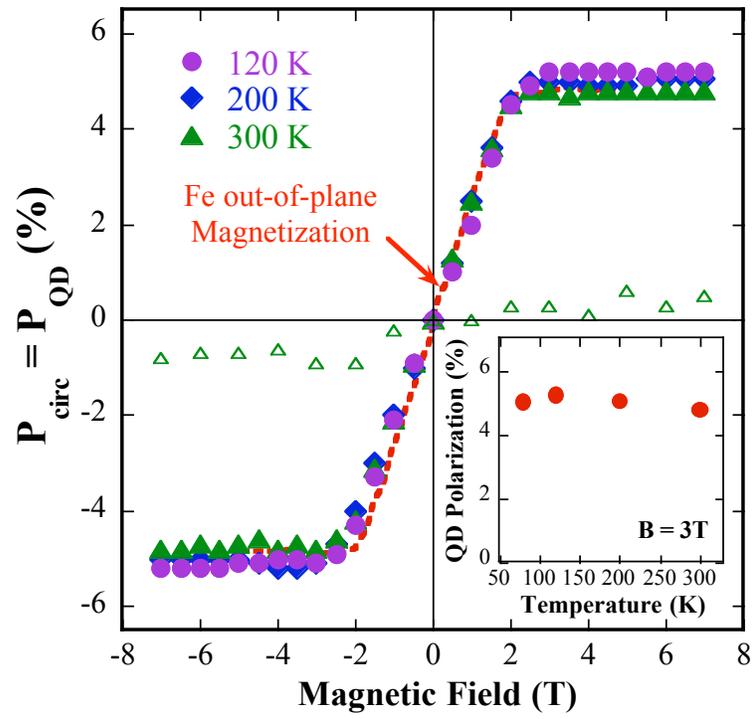

Li *et al*. Fig. 4